\documentclass[12pt]{article}

\usepackage{graphics,graphpap}

\renewcommand{\theequation}{\arabic{equation}}

\newcommand{\bea}{\begin{eqnarray}}
\newcommand{\eea}{\end{eqnarray}}
\newcommand{\beq}{\begin{equation}}
\newcommand{\eeq}{\end{equation}}

\def\msbar{\ifmmode{\overline{\rm MS}} \else{$\overline{\rm MS}$} \fi}
\def\drbar{\ifmmode{\overline{\rm DR}} \else{$\overline{\rm DR}$} \fi}
\def\sf{\ifmmode{\tilde{f}} \else{$\tilde{f}$} \fi}
\def\st{\ifmmode{\tilde{t}} \else{$\tilde{t}$} \fi}
\def\sb{\ifmmode{\tilde{b}} \else{$\tilde{b}$} \fi}
\def\sq{\ifmmode{\tilde{q}} \else{$\tilde{q}$} \fi}
\def\sg{\ifmmode{\tilde{g}} \else{$\tilde{g}$} \fi}
\def\bbar{\ifmmode{\bar{b}} \else{$\bar{b}$} \fi}
\def\tbar{\ifmmode{\bar{t}} \else{$\bar{t}$} \fi}
\def\qbar{\ifmmode{\bar{q}} \else{$\bar{q}$} \fi}
\def\ksla{{k \hspace{-2mm} \slash}}

\newcommand\bvec{\left( \begin{array}{c}}
\newcommand\evec{\end{array}\right)}

\newcommand\bmat{\left( \begin{array}{cc}}
\newcommand\emat{\end{array}\right)}

\newcommand\ch{{\tilde{\chi}}}
\renewcommand\d{\delta}

\def\ksla{{k \hspace{-2.2mm} \slash}}

\newcommand{\sW}{\sin\theta_W}

\newcommand\tw{{\tan{\theta_W}}}

\newcommand\Sgen{{\sum_{\rm gen.}^{}}}

\newcommand\Sf{{\sum_{f=u,\,d}}}

\newcommand\Sa{{\sum_{a=1}^2}}
\newcommand\Sab{{\sum_{a,b=1}^2}}

\newcommand\forpi{{\frac1{\left(4\pi\right)^2}}}

\newcommand\gsh{{\frac{\;g^2}{2}}}

\newcommand{\CL}{{C^f_{L}}}
\newcommand{\CR}{{C^f_{R}}}
\newcommand{\CLs}{{\left(C^f_{L}\right)^2}}
\newcommand{\CRs}{{\left(C^f_{R}\right)^2}}
\newcommand{\IL}{{I_f^{3L}}}

\renewcommand\d{\delta}

\def\su{\ifmmode{\tilde{u}} \else{$\tilde{u}$} \fi}
\def\sd{\ifmmode{\tilde{d}} \else{$\tilde{d}$} \fi}

\begin{document}

\pagestyle{empty} \vspace*{-1cm}
\begin{flushright}
  HEPHY-PUB 738/01 \\
  TU-615 \\
  hep-ph/0104109
\end{flushright}

\vspace*{1.4cm}

\begin{center}
\begin{Large} \bf
One--loop corrections to the chargino and neutralino mass matrices
in the on--shell scheme
\end{Large}

\vspace{10mm}

{\large H. Eberl$^a$, M. Kincel$^{a,\,b}$, W. Majerotto$^a$,
 Y.~Yamada$^c$}

\vspace{6mm}
\begin{tabular}{l}
 $^a${\it Institut f\"ur Hochenergiephysik der \"Osterreichischen
 Akademie der Wissenschaften,}\\
 \hphantom{$^a$}{\it A--1050 Vienna, Austria}\\
 $^b${\it Department of Theoretical Physics FMFI UK, Comenius
 University,  SK-84248}\\
  \hphantom{$^b$}{\it Bratislava, Slovakia }\\
 $^c${\it Department of Physics, Tohoku University,
Sendai 980--8578, Japan}
\end{tabular}

\vspace{20mm}

\begin{abstract}
We present a consistent procedure for the calculation of the
one--loop corrections to the charginos and neutralinos by using
their {\it on--shell} mass matrices. The on--shell gaugino mass
parameters $M$ and $M'$, and the Higgsino mass parameter $\mu$ are
defined by the elements of these on--shell mass matrices. The
on--shell mass matrices are different by finite one--loop
corrections from the tree--level ones given in terms of the
on--shell parameters. When the on--shell $M$ and $\mu$ are
determined by the chargino sector, the neutralino masses receive
corrections up to 4\%. This must be taken into account in
precision measurements at future $e^+ e^-$ linear colliders.
\end{abstract}
\end{center}

\vfill

\newpage
\pagestyle{plain} \setcounter{page}{2}

\section{Introduction}

For the comparison of the Standard Model predictions with the
precision experiments at LEP, calculations at tree--level were no
more sufficient making the inclusion of radiative corrections
necessary. They provided important information on the top quark
mass, the Higgs boson mass and the unification condition for the
gauge couplings.

The future generation of colliders, the Tevatron, LHC and a
$e^+e^-$ linear collider will explore an energy range, where one
expects the appearance of supersymmetric particles. It will be
possible to measure cross sections, branching ratios, masses, etc.
At a $e^+e^-$ linear collider it will even be possible to perform
precision experiments for the production and decay of SUSY
particles \cite{tesla}. This allows one to test the underlying
SUSY model. For instance, for the mass determination of charginos
and neutralinos an accuracy of $\Delta
m_{\ch^{\pm,0}}=0.1$--$1$~GeV might be reached by performing
threshold scans \cite{martyn}. The precision measurements of their
couplings will also be possible
\cite{tesla,peskin,nojiri,choi,mpick}. Therefore, for a comparison
with experiment higher order effects have to be included in the
theoretical calculations.

For calculations of radiative corrections to the masses,
production cross sections, decay rates, etc. of charginos and
neutralinos a proper renormalization of the chargino and
neutralino mixing matrices is needed. One--loop corrections to the
masses were given in \cite{pierce,lahanas} in the scale dependent
\drbar scheme. Chargino production at one--loop was treated by
several authors \cite{diaz,yamada,blank-hollik}. In \cite{diaz}
the \drbar scheme was used where the scale dependence was
cancelled by replacing the tree--level parameters by running ones.
In \cite{yamada} effective chargino mixing matrices were
introduced, which are independent of the renormalization
scale~$Q$. A pure on--shell renormalization scheme was adopted for
the full one--loop radiative corrections to chargino and
neutralino production in $e^+ e^-$ annihilation
\cite{blank-hollik} and squark decays into charginos and
neutralinos \cite{guasch}. In this article we study another type
of correction which has not been discussed so far.

We present a consistent procedure for the calculation of the
one--loop corrections to the on--shell mass matrix of charginos,
$X$, and that of neutralinos, $Y$. One has to distinguish three
types of the mass matrix: the tree--level mass matrix $\tilde X$
$(\tilde Y)$ given in terms of the on--shell parameters, the
\drbar running tree--level matrix $X^0$ $(Y^0)$, and the on--shell
mass matrix $X$ ($Y$) which generates physical (pole) masses and
on--shell mixing matrices by diagonalization. The on--shell
parameters in $\tilde X$ and $\tilde Y$ are given as follows: The
SU(2) gaugino mass parameter $M$ and the Higgsino mass parameter
$\mu$ are defined by the elements of the chargino on--shell mass
matrix $X$, the U(1) gaugino mass parameter $M'$ by the neutralino
mass matrix $Y$, and the other parameters ($m_W$, $m_Z$,
$\sin^2\theta_W$, $\tan\beta$) are given by the gauge and Higgs
boson sectors. We calculate the finite shifts $\Delta
X=X-\tilde{X}$ and $\Delta Y=Y-\tilde{Y}$. Especially, the zero
elements of the tree--level matrix $\tilde{Y}$ receive non--zero
corrections by $\Delta Y$. In the numerical analysis we calculate
the contributions of the fermion and sfermion loops, which are
usually most important. We also discuss the case where the
on--shell $M'$ is defined by the SUSY GUT relation
$M'=\frac53\tan^2{\theta_W}M$.

First we illustrate in Section~\ref{sec:ferfields} our
renormalization procedure for the fermion field with $n$
components. In Section~\ref{sec:charneu} we give the explicit
one--loop corrections to the mass matrices of charginos and
neutralinos.  We work out the shifts in the matrix elements, with
the discussion of the on--shell renormalization of $M$, $\mu$, and
$M'$. The case where the on--shell $M'$ is defined by $M$ is also
discussed. In Section~\ref{sec:numerics} we present some numerical
results of the corrections by fermion and sfermion loops, mainly
the correction to the masses for fixed on--shell parameters.
Conclusions are given in Section~\ref{sec:concl}.

\section{On-shell
renormalization of fermions \label{sec:ferfields}}

In this section we show the on--shell renormalization of the
fermion field $\psi$ with $n$ components (each component being a
four component Dirac spinor) and its mass matrix, following the
formulas in \cite{kniehl,newYamada}.

The mass term of the fermion in the interaction basis is
\begin{equation}
 \label{eq:Vferm1}
V = \bar \psi_R\, M\, \psi_L + \bar \psi_L\, M^\dagger\, \psi_R\,.
\end{equation}
$M$ is a $n\times n$ mass matrix, which is real assuming CP
conservation. With the rotations
\begin{eqnarray}
  \label{eq:psitransR}
f_R &=& U\,\psi_R\, ,\\
  \label{eq:psitransL}
f_L &=& V\,\psi_L\, ,
\end{eqnarray}
the mass matrix can be diagonalized:
\begin{equation}
V = \bar f\, M_D\, f =  \bar f_L\, M_D\, f_R + \bar f_R\, M_D\,
f_L = \sum_{i=1}^n\, m_{f_i}\,  \bar f_{i}\, f_{i}\, ,
\end{equation}
with the diagonal matrix $M_D = \mbox{diag}(m_{f_1},m_{f_2},
\ldots ,m_{f_n})$. For Majorana fermions we allow negative
$m_{f_i}$ in order to keep $U=V$ real. $M_D$ is related to $M$ by:
\begin{equation}
M_D = U\, M \, V^T\, . \label{eq:diag1}
\end{equation}
We express the bare quantities (with superscript 0) by the
renormalized ones:
\begin{eqnarray}
  \label{eq:fLR0}
f_{L,R}^0 & = & ( 1 + \frac{1}{2}\, \delta Z_{L,R})\,f_{L,R} \, ,
\\
  \label{eq:barfLR0}
\bar f_{L,R}^0 & = & \bar f_{L,R}\,( 1 + \frac{1}{2}\, \delta
Z_{L,R}^\dagger) \, , \\
  \label{eq:U0}
U^0 & = & U + \delta U\, , \\
  \label{eq:V0}
V^0 & = & V + \delta V\, , \\
  \label{eq:M0}
M^0 & = & M + \delta M\, .
\end{eqnarray}
$ \delta Z$, $ \delta U$, $ \delta V$, and $ \delta M$ are $(n
\times n)$ matrices. Hence
\begin{equation}
  \label{eq:fR0trans1}
\psi_R^0 = \left(U^T( 1 +  \frac{1}{2}\,\delta Z_{R}) + \delta U^T
\right)\, f_R \, ,
\end{equation}
and
\begin{equation}
  \label{eq:fL0trans1}
\psi_L^0 = \left(V^T( 1 +  \frac{1}{2}\,\delta Z_{L}) + \delta V^T
\right)\, f_L \, .
\end{equation}
By demanding that the counter terms $\delta U$ and $\delta V$
cancel the antisymmetric parts of the wave function corrections,
we get the fixing conditions for $\delta U$~and~$\delta V$:
\begin{eqnarray}
\label{eq:dUgen} \delta U & = & \frac{1}{4}\,(\delta Z_R - \delta
Z_R^T)\, U\,,\\ \label{eq:dVgen} \delta V & = &
\frac{1}{4}\,(\delta Z_L - \delta Z_L^T)\, V\,.
\end{eqnarray}
This is equivalent to redefining the wave function shifts in a
symmetric way,
\begin{eqnarray}
  \label{eq:fR0redef}
f_{L,R}^0 & = & \left(1 +
    \frac{1}{4}(\delta Z_{L,R} + \delta Z_{L,R}^\dagger)\right)
\, f_{L,R} \,, \\ \bar f_{L,R}^{0} & = & \bar f_{L,R} \left( 1 +
 \frac{1}{4}(\delta Z_{L,R}^\dagger + \delta Z_{L,R})\right)\,,
\end{eqnarray}
and setting $\delta U = \delta V = 0$. The renormalization
conditions eqs.~(\ref{eq:dUgen}) and (\ref{eq:dVgen}) have already
been used in \cite{kniehl,blank-hollik}. According to
eq.~(\ref{eq:diag1}) the on--shell mass matrix $M$ is composed by
the on--shell mixing matrices ($U$, $V$) and the pole masses, $M_D
= \mbox{diag}( m_{f_i}(\mbox{\footnotesize pole}))$.\\

We start from the most general form of the matrix element of the
one--loop renormalized two point function for mixing fermions,
\begin{equation}\label{eq:hatGamma}
i\,\hat{\Gamma}_{ij}(k^2) =
 i\left[\d_{ij}\left(\ksla - m_{f_i}\right) +
 \ksla\Big(P_L\hat{\Pi}^L_{ij}(k^2)+P_R\hat{\Pi}^R_{ij}(k^2)\Big)
 \,+\,
 \hat{\Pi}^{S,L}_{ij}(k^2)P_L \,+\,
 \hat{\Pi}^{S,R}_{ij}(k^2)P_R \right]
 \,.
\end{equation}
$\hat\Pi^L$, $\hat\Pi^R$, $\hat\Pi^{S,L}$, and $\hat\Pi^{S,R}$ are
the fermion self--energy matrices. The "hat" denotes the
renormalized quantities. Then the mass shifts $\delta m_{f_k}$ are
given by:
\begin{equation}
\delta m_{f_k} =  \frac12\,{\rm Re}\Bigg[
 m_{f_k}\, \Big(
  \Pi^L_{kk}( m^2_{f_k})+
  \Pi^R_{kk}( m^2_{f_k})
  \Big) +
 \Pi^{S,L}_{kk}( m^2_{f_k})+
 \Pi^{S,R}_{kk}( m^2_{f_k})
 \Bigg]
 \label{eq:dmfk} \, ,
\end{equation}
and the off--diagonal wave function renormalization constants of
$\delta Z_R^{}$ and $\delta Z_L^{}$ read ($i \ne j$):
\begin{eqnarray}
\left(\delta Z_R^{}\right)_{ij} &=&
 \frac{2}{ m^2_{f_i} -
m^2_{f_j}} \, {\rm Re}\Bigg[
 \Pi^R_{ij}( m^2_{f_j}) m^2_{f_j} +
 \Pi^L_{ij}( m^2_{f_j}) m_{f_i} m_{f_j}
 \nonumber\\ && \hspace{4cm}
+\, \Pi^{S,R}_{ij}( m^2_{f_j}) m_{f_i}
 + \Pi^{S,L}_{ij}( m^2_{f_j}) m_{f_j}  \Bigg]
\, . \label{eq:dZRchij}
\end{eqnarray}
$\left(\delta Z_L^{}\right)_{ij}$ is obtained by replacing $L\,
\leftrightarrow\, R$ in eq.~(\ref{eq:dZRchij}).
The counterterm for the mass matrix element $\delta M_{ij}$ can be
written as:
\begin{equation}
\delta M_{ij} =  \frac12\,\sum_{k,l} U_{ki}V_{lj}{\rm Re}\Big[
  \Pi^L_{kl}( m^2_{f_k})m_{f_k} +
  \Pi^R_{kl}( m^2_{f_l})m_{f_l} +
 \Pi^{S,L}_{kl}( m^2_{f_k})+
 \Pi^{S,R}_{lk}( m^2_{f_l})
 \Big]
\, . \label{eq:dMij}
\end{equation}


\pagebreak

\section{Chargino and neutralino mass matrices\\
 at one--loop level}
\label{sec:charneu}

In the MSSM the chargino mass matrix is given by:
 \begin{equation}
        X = \bmat M & \sqrt{2}\, m_W \sin\beta \\
                 \sqrt{2}\, m_W \cos\beta & \mu \emat \, .
  \label{eq:chi+mat}
\end{equation}
It is diagonalized by the two real $(2 \times 2)$ matrices $U$ and
$V$:
\begin{equation}
        U\, X\, V^T = M_D = \bmat
m_{\tilde\chi^+_1} & 0 \\ 0 & m_{\tilde\chi^+_2} \emat \, ,
 \label{eq:UVchi}
\end{equation}
with $m_{\tilde\chi^+_1}$, and $m_{\tilde\chi^+_2}$ the physical
masses of the charginos (choosing $m_{\tilde\chi^+_1} <
m_{\tilde\chi^+_2}$). The shifts for $U$ and $V$ are then given by
eqs.~(\ref{eq:dUgen}) and (\ref{eq:dVgen}):
\begin{eqnarray}
\label{eq:dUch} \delta U & = & \frac{1}{4}\,(\delta
Z_R^{\tilde\chi^+}
 - \delta Z_R^{\tilde\chi^+\,T})\, U\,,\\
 \label{eq:dVch}
\delta V & = & \frac{1}{4}\,(\delta Z_L^{\tilde\chi^+}
 - \delta Z_L^{\tilde\chi^+\,T})\, V\, .
\end{eqnarray}
The shift in $X$ follows from
\begin{equation}
 \delta X =  \delta (U^T\, M_D\, V) = \delta U^T\, M_D\, V + U^T\, M_D\, \delta V +
U^T\, \delta M_D\, V \, . \label{eq:delX}
\end{equation}
Its matrix elements are:
\begin{eqnarray}
 \left(\delta X\right)_{ij} & = &
 \sum_{k=1}^2 \left[ m_{\tilde\chi^+_k} \left(
\delta U_{ki} V_{kj} +  U_{ki} \delta V_{kj}\right) + \delta
m_{\tilde\chi^+_k} U_{ki} V_{kj}\right]\, , \label{eq:dXij}
\end{eqnarray}
where the elements $\delta U_{ki}$ and $\delta V_{kj}$ are
obtained from eqs.~(\ref{eq:dUch}) and (\ref{eq:dVch}) together
with eq.~(\ref{eq:dZRchij}). $\delta m_{\tilde\chi^+_k}$ is given
by eq.~(\ref{eq:dmfk}). The explicit forms for the chargino
self--energies are given in eq.~(\ref{eq:chargSE}).

Now we want to calculate the on--shell mass matrix $X$ at
one--loop level. We first show the relation between three types of
the mass matrix, $X$, $\tilde X$, and $X^0$. $\tilde X$ is the
tree--level mass matrix, which has the form of
eq.~(\ref{eq:chi+mat}) in terms of the on--shell parameters ($M$,
$\mu$, $m_W$, $\tan\beta$). $\tilde X$ is diagonalized by the
matrices $\tilde U$ and $\tilde V$ to give the eigenvalues $\tilde
m_i$. The bare mass matrix (or the \drbar running tree--level
matrix) $X^0$ is related to $\tilde X$ by
\begin{equation}\label{eq:X0def}
 X^0 = \tilde X + \delta_c X\, ,
\end{equation}
where $\delta_c$ means the variation of the (on--shell) parameters
in $\tilde X$. The correction (\ref{eq:delX}) represents the
difference between $X^0$ and the on--shell $X$, with $X^0 = X+
\delta X$. This implies
\begin{equation}\label{eq:Xdef}
 X = \tilde X + \delta_c X -  \delta X = \tilde X + \Delta X\, .
\end{equation}
The one--loop corrected matrix $X$ is the sum of the tree--level
mass matrix $\tilde X$ with the on--shell quantities and the
ultraviolet (UV) finite shifts $\Delta X$.

To discuss the shifts $\Delta X$ we need to fix the definition of
the on--shell parameters in $\tilde X$. In this article we define
the on--shell parameters $M$ and $\mu$ by the elements of the
on--shell mass matrix of charginos, by $M = X_{11}$ and $\mu =
X_{22}$, respectively. This definition gives the counterterms
\begin{eqnarray}
\delta M & = &  \left(\delta X\right)_{11}\, ,\\
 \delta \mu & = & \left(\delta X\right)_{22}\, .
\end{eqnarray}
We later comment on the case where $M$ and $\mu$ are fixed by the
neutralino mass matrix. In addition, we fix the on--shell $m_W$ as
the physical (pole) mass and $\tan\beta$ by the condition in the
Higgs sector, as given in the Appendix. As a result we have:
\begin{eqnarray}
  \label{eq:DeltaX11}
\Delta X_{11} &=& 0\,\\[2mm]
  \label{eq:DeltaX12}
\Delta X_{12} &=& \left(\frac{\delta m_W}{m_W} + \cos^2\beta\,
\frac{\delta \tan\beta}{\tan\beta} \right)\,
 X_{12} - \delta X_{12}\,\\[2mm]
  \label{eq:DeltaX21}
\Delta X_{21} &=& \left(\frac{\delta m_W}{m_W} - \sin^2\beta\,
\frac{\delta \tan\beta}{\tan\beta} \right)\,
 X_{21} - \delta X_{21}\,\\[2mm]
  \label{eq:DeltaX22}
\Delta X_{22} &=& 0\, ,
\end{eqnarray}
with $\delta X_{12}$ and $\delta X_{21}$ given by
eq.~(\ref{eq:delX}) with the replacements $U \to \tilde U$, $V \to
\tilde V$, $m_{\tilde \chi_k} \to \tilde m_k$. The counterterm
$\delta m_W$ is given in eq.~(\ref{eq:dmV}) together with
eq.~(\ref{eq:Wself}). $\delta \tan\beta$ is obtained from
eqs.~(\ref{eq:dtanb}) and (\ref{eq:A0Z0self}). By diagonalizing
the matrix $X$ one gets the one--loop pole masses of charginos,
$m_{\tilde\chi^+_{1,2}}$, and their on--shell rotation matrices
$U$ and $V$, which enter in all chargino couplings.

If the chargino masses $m_{\tilde\chi^+_{1,2}}$ are known from
experiment (e.~g. from a threshold scan), one first calculates the
tree--level parameters $\tilde M$, $\tilde \mu$, $\tilde U$, and
$\tilde V$, using eqs.~(\ref{eq:chi+mat}, \ref{eq:UVchi}) together
with the experimental information of chargino
couplings~\cite{choi,mpick,kneur}. $\delta \tilde U$ and
$\delta\tilde V$ are then obtained from eqs.~(\ref{eq:dUch}) and
(\ref{eq:dVch}), depending on the sfermion parameters. This
enables one to calculate $\Delta X_{12}$ and $\Delta X_{21}$ and
the one--loop corrected mass matrix $X$. By requiring that $X$
give the measured chargino masses $m_{\tilde\chi^+_{1,2}}$, one
then gets the correct on--shell parameters $M$ and $\mu$. The
error that one starts from $\tilde M$ and $\tilde \mu$ is of
higher order. The dependence of this procedure on sfermion
parameters will be discussed in Section 4.

Let us now turn to the neutralino sector. The mass matrix in the
interaction basis has the form
\begin{equation}
        Y = \left(\begin{array}{cccc}
M' & 0 & - m_Z \sin\theta_W \cos\beta & m_Z \sin\theta_W
\sin\beta\\ 0 & M & m_Z \cos\theta_W \cos\beta & -m_Z \cos\theta_W
\sin\beta\\ - m_Z \sin\theta_W \cos\beta & m_Z \cos\theta_W
\cos\beta & 0 & -\mu\\ m_Z \sin\theta_W \sin\beta & -m_Z
\cos\theta_W \sin\beta & -\mu & 0
\end{array}\right) \, .
 \label{neumat1}
\end{equation}
Since we assume CP conservation this matrix is real and symmetric.
It is diagonalized by the real matrix~$Z$:
\begin{equation}
        Z\, Y\, Z^T = M_D = \mbox{diag}(m_{\tilde\chi_1^0},
 \,m_{\tilde\chi_2^0}, \,m_{\tilde\chi_3^0},\,m_{\tilde\chi_4^0}) \,
 .
 \label{neudiag1}
\end{equation}
We allow negative values for the mass parameters
$m_{\tilde\chi_i^0}$. The convention $|m_{\tilde\chi_1^0}| <
|m_{\tilde\chi_2^0}| < |m_{\tilde\chi_3^0}| <
|m_{\tilde\chi_4^0}|$ is used.
 \\
The shift for the rotation matrix $Z$ is obtained from
eqs.~(\ref{eq:dUgen}) and (\ref{eq:dVgen}) by the substitutions $U
\to Z$ and $V \to Z$,
\begin{eqnarray}
\delta Z & = & \frac{1}{4}\,(\delta Z_L^{\tilde\chi^0}
 - \delta Z_L^{\tilde\chi^0\,T})\, Z,\\
\delta Z^T & = & \frac{1}{4}\,Z^T (\delta Z_R^{\tilde\chi^0\,T} -
\delta Z_R^{\tilde\chi^0}) \, .
\end{eqnarray}
Note that  $\delta Z_L^{\tilde\chi^0} = \delta Z_R^{\tilde\chi^0}$
due to the Majorana character of the neutralinos. The shift
$\delta Y$ is given by $\delta Y = \delta (Z^T M_D Z)$, i. e. for
the matrix elements
\begin{eqnarray}
 (\delta Y)_{ij} & = &
 \sum_{k=1}^4 \left[
\delta m_{\tilde \chi^0_k} Z_{ki} Z_{kj} +   m_{\tilde \chi^0_k}
\delta Z_{ki} Z_{kj} + m_{\tilde \chi^0_k} Z_{ki} \delta
Z_{kj}\right]\, . \label{eq:dYij2}
\end{eqnarray}
The wave function correction terms $\delta Z^{\tilde\chi^0}$ are
given by eq.~(\ref{eq:dZRchij}) and $\delta m_{\tilde \chi^0_k}$
by eq.~(\ref{eq:dmfk}). The formulas for the neutralino
self--energies are shown in eqs.~(\ref{eq:neuSE}) and
(\ref{eq:neuSES}).

We again start with the tree--level mass matrix $Y^{\rm tree}
\equiv \tilde Y$ which has the form of eq.~(\ref{neumat1}) in
terms of the on--shell parameters ($M$, $\mu$, $M'$, $m_Z$,
$\sin\theta_W$, $\tan\beta$). First one calculates the tree--level
masses $\tilde m_k$ and the rotation matrix $\tilde Z$ by
diagonalizing $\tilde Y$. In analogy to the chargino case, the
one--loop on--shell mass matrix $Y$ is
\begin{equation}
  \label{eq:Ycorr}
Y = Y^0 - \delta Y = \tilde Y + \delta_c Y - \delta Y = \tilde Y +
\Delta Y\, .
\end{equation}
Here $\delta_c$ means the variation of the parameters in $\tilde
Y$. Again $\Delta Y$ is UV finite.

We need to fix the on--shell input parameters in $\tilde Y$. The
on--shell $M$ and $\mu$ are already determined by the chargino
sector. We define the on--shell parameter $M'$ by the on--shell
mass matrix of neutralinos as $Y_{11} = M'$. This condition gives
\begin{equation}\label{eq:Mprimedef}
 \delta M' = (\delta Y)_{11} \, .
\end{equation}
We further fix the on--shell $m_Z$,
$\sin^2\theta_W=1-m_W^2/m_Z^2$, and $\tan\beta$ in the same way as
in the case of charginos.

The 10 independent entries of the real and symmetric matrix
$\Delta Y$ are
\begin{eqnarray}
  \label{eq:DeltaY11}
\Delta Y_{11} &=& 0\,\\[2mm]
  \label{eq:DeltaY12}
\Delta Y_{12} &=& - \delta Y_{12}\,\\[2mm]
  \label{eq:DeltaY13}
\Delta Y_{13} &=& \left(\frac{\delta m_Z}{m_Z}+ \frac{\delta
 \sW}{\sW} - \sin^2\beta\, \frac{\delta \tan\beta}{\tan\beta}
 \right)\, Y_{13} - \delta Y_{13}\,\\[2mm]
  \label{eq:DeltaY14}
\Delta Y_{14} &=& \left(\frac{\delta m_Z}{m_Z}+ \frac{\delta
 \sW}{\sW} + \cos^2\beta\, \frac{\delta \tan\beta}{\tan\beta}
 \right)\, Y_{14} - \delta Y_{14}\,\\[2mm]
  \label{eq:DeltaY22}
\Delta Y_{22} &=& \delta M - \delta Y_{22}\, \,\\[2mm]
  \label{eq:DeltaY23}
\Delta Y_{23} &=& \left(\frac{\delta m_Z}{m_Z} - \tan^2\theta_W\,
\frac{\delta \sW}{\sW} - \sin^2\beta\, \frac{\delta
\tan\beta}{\tan\beta}
 \right)\, Y_{23} - \delta Y_{23}\,\\[2mm]
  \label{eq:DeltaY24}
\Delta Y_{24} &=& \left(\frac{\delta m_Z}{m_Z} - \tan^2\theta_W\,
\frac{\delta \sW}{\sW} + \cos^2\beta\, \frac{\delta
\tan\beta}{\tan\beta}
 \right)\, Y_{24} - \delta Y_{24}\,\\[2mm]
  \label{eq:DeltaY33}
\Delta Y_{33} &=&  - \delta Y_{33}\,\\[2mm]
  \label{eq:DeltaY34}
\Delta Y_{34} &=& -\delta\mu - \delta Y_{34}\,\\[2mm]
  \label{eq:DeltaY44}
\Delta Y_{44} &=& - \delta Y_{44}\, ,
\end{eqnarray}
with $\delta Y_{ij}$ given by eq.~(\ref{eq:dYij2}) with $Z \to
\tilde Z$, and $m_{\tilde \chi_k^0} \to \tilde m_{\tilde
\chi_k^0}$. $\delta \sW$ and $\delta m_Z$ can be calculated from
eqs.~(\ref{eq:dmV}) -- (\ref{eq:CLR}). Notice that the elements
$Y_{12}=Y_{21}$, $Y_{33}$, and $Y_{44}$ are no more zero. Recall
that $\delta M = \delta X_{11}$, and $\delta \mu = \delta X_{22}$.
The corrected neutralino masses and the corrected rotation matrix
$Z$ are obtained by diagonalizing the matrix $Y$,
eq.~(\ref{eq:Ycorr}).

So far we have treated $M'$ as an independent parameter to be
determined in the neutralino sector. If we assume a relation
between gaugino masses, we may define the on--shell $M'$ as a
function of other on--shell parameters instead of $Y_{11}$. The
shift $\Delta Y_{11}$ is then no longer zero. For example, when
the SUSY SU(5) relation $M'=\frac{5}{3}\tan^2\theta_WM$ holds for
the \drbar parameters, and the on--shell $M'$ is defined by
imposing the same relation on the on--shell parameters, one has
 \begin{equation}
  \label{eq:DeltaY11Uni}
 \Delta Y_{11}
 \;=\;\bigg(\frac2{\cos^2{\theta_W}}\,\frac{\delta\sW}\sW \,+\,
 \frac{\delta M}{M}\bigg)\,Y_{11}\;-\;\delta Y_{11}\, .
 \end{equation}
We note that eq.~(\ref{eq:DeltaY11Uni}) is also applicable in
other models for gaugino masses, e.g. in the anomaly mediated SUSY
breaking model \cite{AMSB,feng} where $M'=11\tan^2\theta_WM$.

Finally, we would like to remark that one could also first
determine the on--shell values of $M'$, $M$, $\mu$ from the
neutralino sector, that means  $\Delta Y_{11}\,=\,\Delta
Y_{22}\,=\,\Delta Y_{34}\,=\,\Delta Y_{43}\,=\,0$, see
eq.~(\ref{neumat1}) and (\ref{eq:DeltaY11}--\ref{eq:DeltaY44}).
This would imply corrections $\Delta X_{11}$ and $\Delta X_{22}$
in the chargino system.


\section{Numerical examples}
\label{sec:numerics}

In this section we will give some numerical examples for the
on--shell one--loop corrected mass matrices and masses of the
charginos and neutralinos. We take into account the contributions
from all fermions and sfermions.

For simplicity, we will take in the following (if not specified
otherwise) for the soft breaking sfermion mass parameters of the
first and second generation $M_{\tilde Q_{1,2}}=M_{\tilde
U_{1,2}}=M_{\tilde D_{1,2}}=M_{\tilde L_{1,2}}=M_{\tilde
E_{1,2}}$, of the third generation $M_{\tilde
Q_{3}}=\frac{10}9M_{\tilde U_{3}}=\frac{10}{11}M_{\tilde
D_{3}}=M_{\tilde L_{3}}=M_{\tilde E_{3}} = M_{\tilde Q}$, and for
the trilinear couplings $A_t=A_b=A_{\tau}=A$. We take
$m_t=175$~GeV, $m_b=5$~GeV, $m_Z=91.2$~GeV, $m_W=80$~GeV, and
$m_{A^0} = 500$~GeV. Thus the input parameter set is
$\{\,\tan{\beta},M_{\tilde Q_1},\,M_{\tilde
Q},\,A,\,M,\,M',\,\mu\}$.

We always assume that the (on--shell) values of $M$ and $\mu$ are
obtained from the chargino sector as described in Section 3. Then
the chargino mass matrix only gets corrections in the
off--diagonal elements of the matrix $X$. In general, the
corrections to the chargino masses are small ($<1\%$). For
instance, for $\tan{\beta}=7$ and $\{M_{\tilde Q_1},\,M_{\tilde
Q},\,A,\,M,\,\mu\}= \{300,300,-500,300,-400\}$~GeV one gets for
$\Delta X_{12}/X_{12}\simeq0.7/112$ and for \linebreak $\Delta
X_{21}/X_{21}\simeq\nolinebreak-1.1/16$, $\Delta
m_{\tilde\chi^+_1}/m_{\tilde\chi^+_1}=-0.24\,\%$ and $\Delta
m_{\tilde\chi^+_2}/m_{\tilde\chi^+_2}=-0.14\,\%$. For the same
parameters but $\tan{\beta}=40$ one gets for $\Delta
X_{12}/X_{12}\simeq2.3/113$, $\Delta X_{21}/X_{21}\simeq-3.2/2.8$,
$\Delta m_{\tilde\chi^+_1}/m_{\tilde\chi^+_1}=-0.76\,\%$ and
$\Delta m_{\tilde\chi^+_2}/m_{\tilde\chi^+_2}=-0.46\,\%$.

\begin{figure}[h!]
 \begin{center}
 \hspace{-10mm}
  \mbox{\resizebox{85mm}{!}{\includegraphics{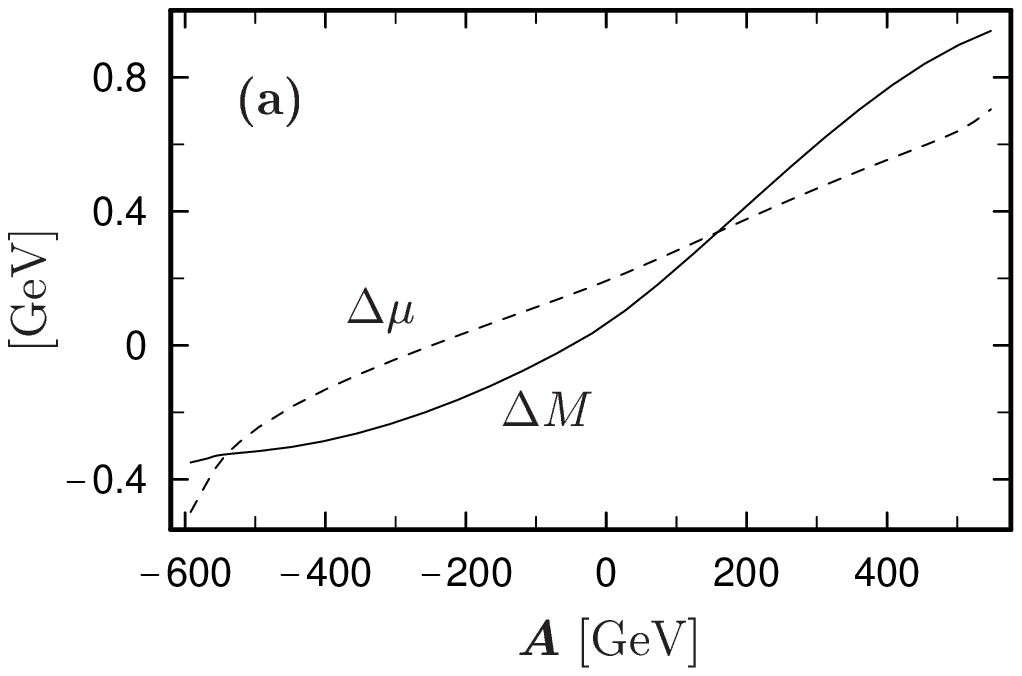}}}
  \hspace{-6mm}
 \mbox{\resizebox{85mm}{!}{\includegraphics{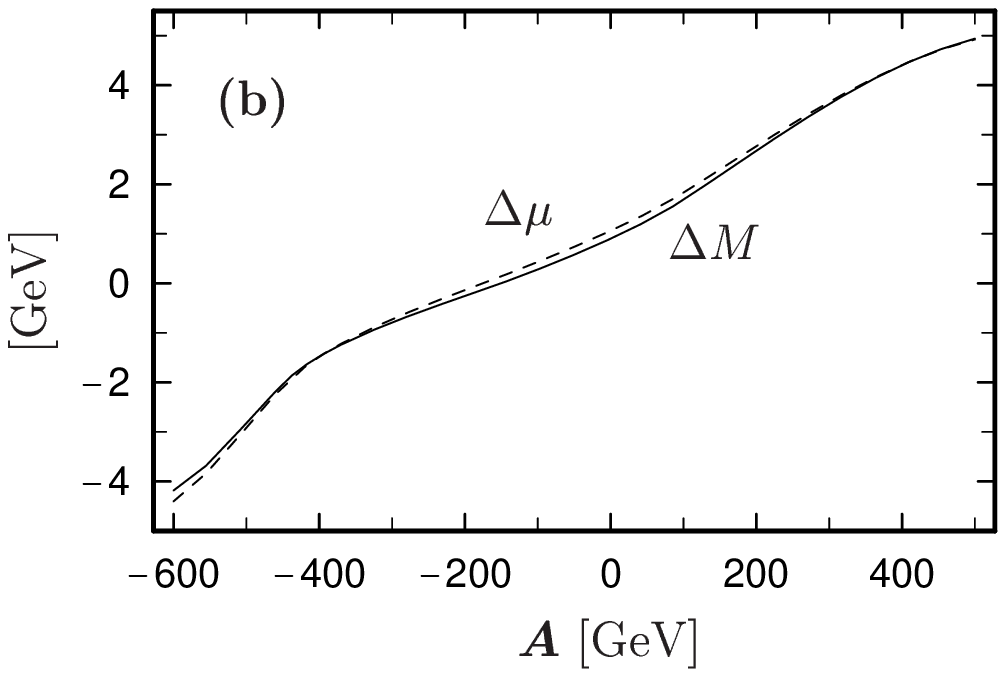}}}
 \hspace{-10mm}
  \vspace{-2mm}
 \caption[fig0]
 {$\Delta M = M - M^{\rm eff.}$ (full lines) and $\Delta \mu = \mu - \mu^{\rm eff.}$
 (dashed lines) as functions of $A$ with $\tan{\beta}=7$, $M_{\tilde
Q_1} = M_{\tilde Q}= 300$~GeV, and ${\rm sign}(\mu) = -1$. In (a)
\{$m_{\tilde \chi_1^+},\, m_{\tilde \chi_2^+}\}$ = \{126.7,
322\}~GeV giving \{$M^{\rm eff.}$, $\mu^{\rm eff.}$\} = \{300,
$-$130\}~GeV for $M > |\mu|$ . In (b) \{$m_{\tilde \chi_1^+},\,
m_{\tilde \chi_2^+}\}$ = \{200, 300\}~GeV giving \{$M^{\rm eff.}$,
$\mu^{\rm eff.}$\} = \{227.7, $-$255.6\}~GeV for $M < |\mu|$.
 \label{fig:0}
 }
 \end{center}
 \end{figure}

As shown in Section 3, the on--shell parameters $M$ and $\mu$ for
given values of the pole masses $m_{\tilde\chi^+_1}$ and
$m_{\tilde\chi^+_2}$ depend on sfermion parameters. In
Fig.~\ref{fig:0} we show the values of $M$ and $\mu$ as functions
of $A$, for fixed $m_{\tilde\chi^+_{1,2}}$, $\tan{\beta}=7$, and
$M_{\tilde Q_1} = M_{\tilde Q} = 300$~GeV. For comparison we show
the difference from the effective parameters used in
\cite{blank-hollik}, $M^{\rm eff.}$ and $\mu^{\rm eff.}$. These
are obtained from the pole masses $m_{\tilde\chi^+_{1,2}}$ by
tree--level sum rules and are therefore independent of sfermion
parameters. We see that the dependence on the sfermion parameters
becomes large for $M\sim |\mu|$, i.~e. large gaugino--Higgsino
mixing.

Let us now discuss the neutralino sector for the on--shell $M$ and
$\mu$ fixed by the chargino sector. We first treat the on--shell
$M'$ as an independent parameter. Then the one-loop corrections to
the mass matrix (\ref{neumat1}) are calculated by
eqs.~(\ref{eq:DeltaY11})--(\ref{eq:DeltaY44}).

\begin{figure}[h!]
 \vspace{-5mm}
 \begin{center}
 \hspace{-10mm}
  \mbox{\resizebox{85mm}{!}{\includegraphics{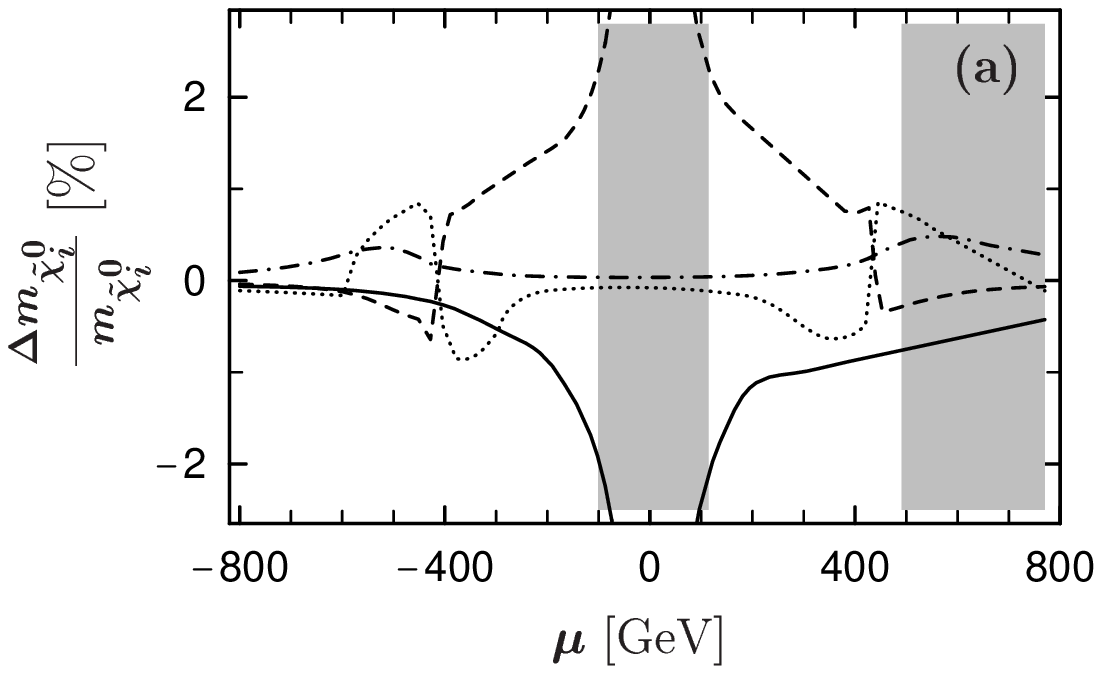}}}
  \hspace{-6mm}
 \mbox{\resizebox{85mm}{!}{\includegraphics{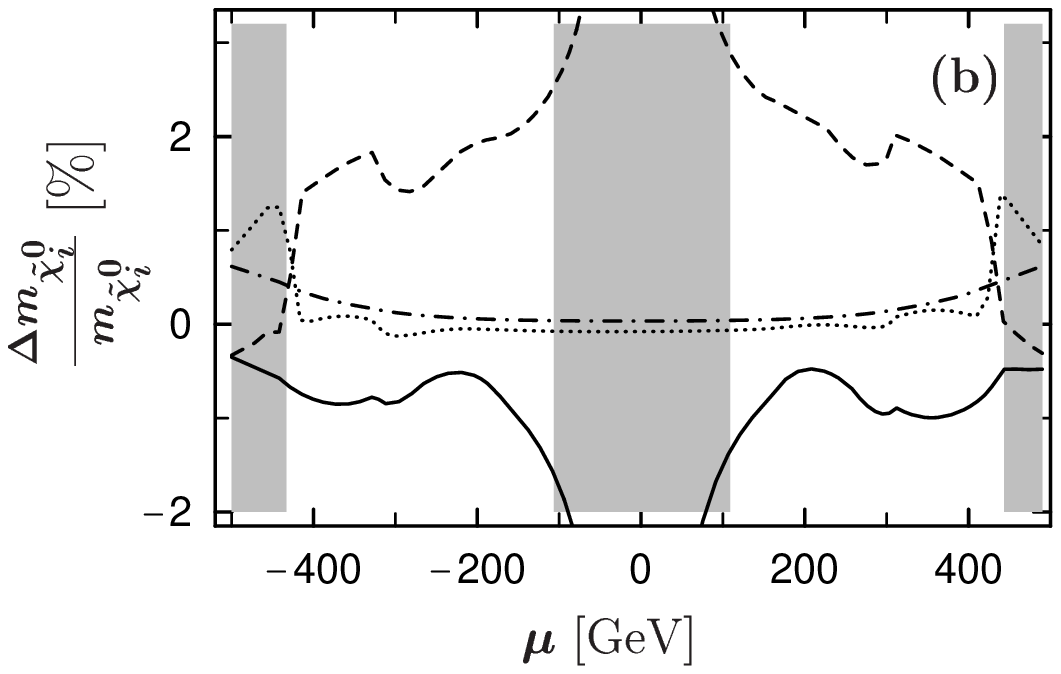}}}
 \hspace{-10mm}
  \vspace{-5mm}
 \caption[fig1]
 {Relative corrections to neutralino masses
 as a function of $\mu$ for $\tan{\beta}=7$ (a) and
$\tan{\beta}=40$ (b) with $\{M_{\tilde Q_1},\,M_{\tilde
Q},\,A,\,M,\,M'\}= \{300,300,-500,300,149.4\}$~GeV.
 \label{fig:mudep}
 The full, dashed, dotted, dash--dotted line corresponds to
 $\ch^0_1$, $\ch^0_2$, $\ch^0_3$, $\ch^0_4$ mass corrections respectively.
 The grey areas are excluded by the bounds $m_{\ch^\pm_1}\geq100$~GeV,
 $m_{h^0}>95$~GeV.}
 \end{center}
 \end{figure}

In Fig.~\ref{fig:mudep}a we show the relative correction $\delta
m_{\tilde\chi^0_i}/m_{\tilde\chi^0_i}$ as a function of $\mu$ for
$\tan{\beta}=7$ and $\{M_{\tilde Q_1},\,M_{\tilde
Q},\,A,\,M,\,M'\}= \{300,300,-500,300,149.4\}$~GeV.

One can see that the corrections to $m_{\tilde\chi^0_1}$ and
$m_{\tilde\chi^0_2}$ can go up to $2.2\%$ for $|\mu|\sim100$~GeV,
where ${\tilde\chi^0_1}$ and ${\tilde\chi^0_2}$ are
higgsino--like. Fig.~\ref{fig:mudep}b shows the same as
Fig.~\ref{fig:mudep}a for $\tan{\beta}=40$ with the other
parameters unchanged. The general behaviour of the curves is very
similar.

We also present numerical values of the mass matrix $\tilde Y$ and
its correction $\Delta Y$ calculated for the same set of
parameters as in Fig.~\ref{fig:mudep}a, with $\mu=110$~GeV.
%
$$ \tilde Y \;+\; \Delta Y \;=\; \left(\begin{array}{cccc}
 149.4 & 0 & -6.2 & 43.3\\
 0 & 300 & 11.3 & -79.2 \\
 -6.2 & 11.3 & 0 & 110 \\
 43.3 & -79.2 & 110 & 0  \\
 \end{array}\right){\rm GeV}
 \;+\;
 \left(\begin{array}{cccc}
 0 & 0.3 & 0.0 & 0.9\\
 0.3 & -0.1 & -0.1 & -0.2 \\
 0.0 & -0.1 & -0.0 & 0.2 \\
 0.9 & -0.2 & 0.2 & -4.3  \\
 \end{array}\right){\rm GeV}\, .
$$
Notice, that also the zero elements of $\tilde Y$ get nonzero
corrections. Especially, the most important contribution comes
from the element $\Delta Y_{44}$, that is from the $\tilde H^0_2$
to $\tilde H^0_2$ transition via a $t\tilde t$ loop. The effects
of $\Delta Y_{22}$, $\Delta Y_{33}$, $\Delta Y_{34}$, $\Delta
Y_{44}$ to the masses were discussed in the limiting cases, for
$|\mu|\ll(M, M')$ in \cite{pomarol} and for $M\ll(|\mu|, M')$ in
\cite{chen,feng}, respectively.

\begin{figure}[h!]
 \begin{center}
 \vspace{-10mm}
 \mbox{\resizebox{85mm}{!}{\includegraphics{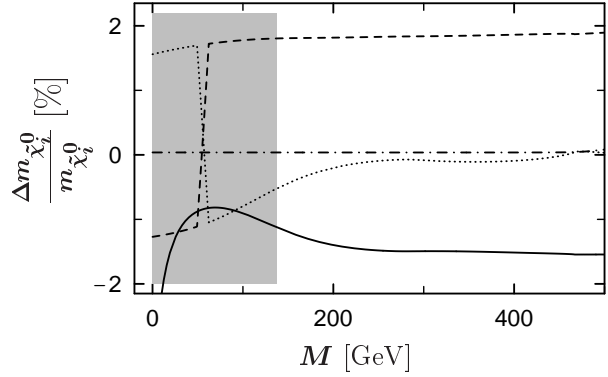}}}
  \vspace{-6mm}
 \end{center}
 \caption[fig2]
  {Relative corrections to neutralino masses
 as a function of $M$ for $\tan{\beta}=7$ and $\{M_{\tilde Q_1},\,M_{\tilde
Q},\,A,\,\mu\}= \{300,300,-500,-130\}$~GeV.
 The full, dashed, dotted, dash--dotted line corresponds to
 $\ch^0_1$, $\ch^0_2$, $\ch^0_3$, $\ch^0_4$ mass corrections respectively.
 The grey areas are excluded by the bound $m_{\ch^\pm_1}\geq100$~GeV.
 \label{fig:Mdep} }
 \end{figure}

In Fig.~\ref{fig:Mdep} we show the $M$--dependence with $M'=
0.498\, M$ for $\mu=-130$~GeV and the other parameters as in
Fig.~\ref{fig:mudep}a. One sees that up to $M\simeq200$~GeV the
$M$--dependence of ${\delta
m_{\tilde\chi^0_1}}/{m_{\tilde\chi^0_1}}$ is rather strong and
becomes weak when $\tilde\chi^0_1$ becomes higgsino--like. One
also sees the various discontinuities in $\tilde\chi^0_2$ and
$\tilde\chi^0_3$ due to 'level crossings' (see e. g.
\cite{majer-fraas}).

 \begin{figure}[h!]
 \begin{center}
  \vspace{-10mm}
 \mbox{\resizebox{85mm}{!}{\includegraphics{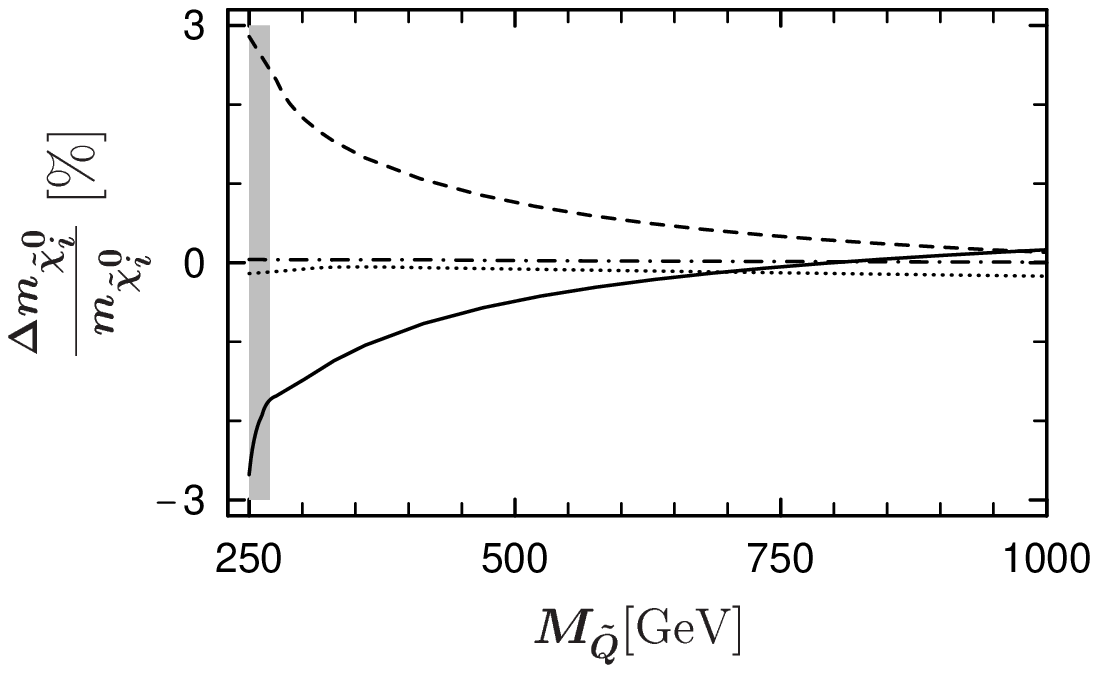}}}
 \end{center}
   \vspace{-10mm}
 \caption[fig3]
  {Relative corrections to neutralino masses
 as a function of $M_{\tilde Q}$ for $\tan{\beta}=7$ and
$\{M_{\tilde Q_1},\,A,\,M,\,M',\,\mu\}= \{300,-500,
300,149.4,-130\}$~GeV.
 The full, dashed, dotted, dash--dotted line corresponds to
 $\ch^0_1$, $\ch^0_2$, $\ch^0_3$, $\ch^0_4$ mass corrections respectively.
 The grey areas are excluded by the bound $m_{\tilde t_1}\geq100$~GeV.
 \label{fig:MsQdep} }
 \end{figure}

Fig.~\ref{fig:MsQdep} exhibits the dependence on $M_{\tilde{Q}}$
for the same parameter set as in Fig.~\ref{fig:mudep}a with
$\mu=-130$~GeV. The corrections to the masses become smaller with
increasing $M_{\tilde{Q}}$. The dependence on $M_{\tilde{Q}_1}$
for fixed $M_{\tilde{Q}}$ is very small. The effects of the
non-decoupling corrections \cite{chankowski,yamada} to the
gaugino-Higgsino mixing elements of $Y$ and $X$ cannot be seen in
Fig.~\ref{fig:MsQdep}.

\begin{figure}[h!]
 \begin{center}
  \vspace{-10mm}
 \mbox{\resizebox{85mm}{!}{\includegraphics{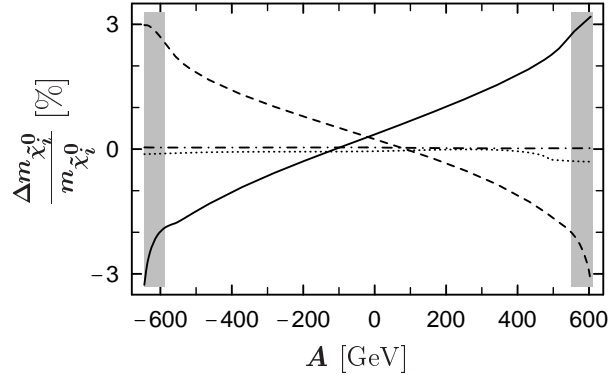}}}
 \end{center}
 \vspace{-10mm}
 \caption[fig4]
   {Relative corrections to neutralino masses
 as a function of $A$ for $\tan{\beta}=7$ and
$\{M_{\tilde Q_1},\,M_{\tilde Q},\,M,\,M',\,\mu\}= \{300,300,
300,149.4,-130\}$~GeV.
 The full, dashed, dotted, dash--dotted line corresponds to
 $\ch^0_1$, $\ch^0_2$, $\ch^0_3$, $\ch^0_4$ mass corrections respectively.
 The grey areas are excluded by the bound $m_{\tilde t_1}\geq100$~GeV.
 \label{fig:Atdep} }
 \end{figure}

In Fig.~\ref{fig:Atdep} we show the dependence on $A_t$, with
$\mu=-130$~GeV, $A_b=A_\tau$ fixed to $500$~GeV  and the other
parameters as in Fig.~\ref{fig:mudep}a. It is strong for the
higgsino--like states $\tilde\chi^0_1$ and $\tilde\chi^0_2$ being
mainly due to the $A_t$ dependence of $\Delta Y_{44}$. The
sensitivity to the value of $A_b$ and $A_\tau$ is very weak.

\begin{figure}[h!]
 \begin{center}
  \vspace*{-10mm}
 \hspace{-5mm}
 \mbox{\resizebox{84mm}{!}{\includegraphics{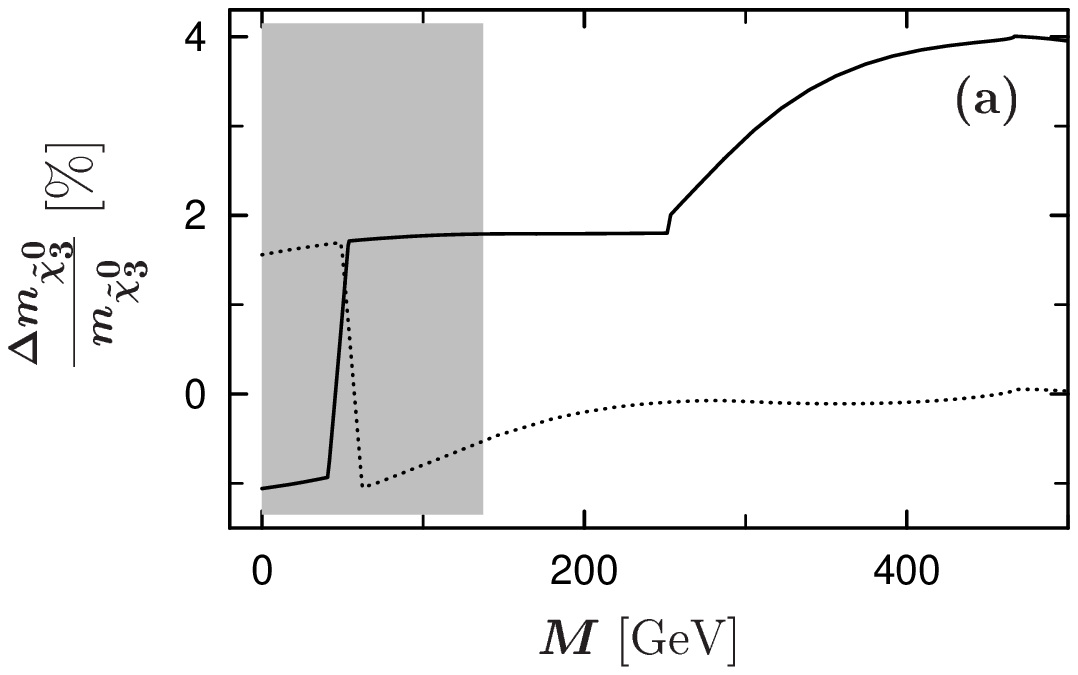}}}
 \hspace{-7mm}
 \mbox{\resizebox{84mm}{!}{\includegraphics{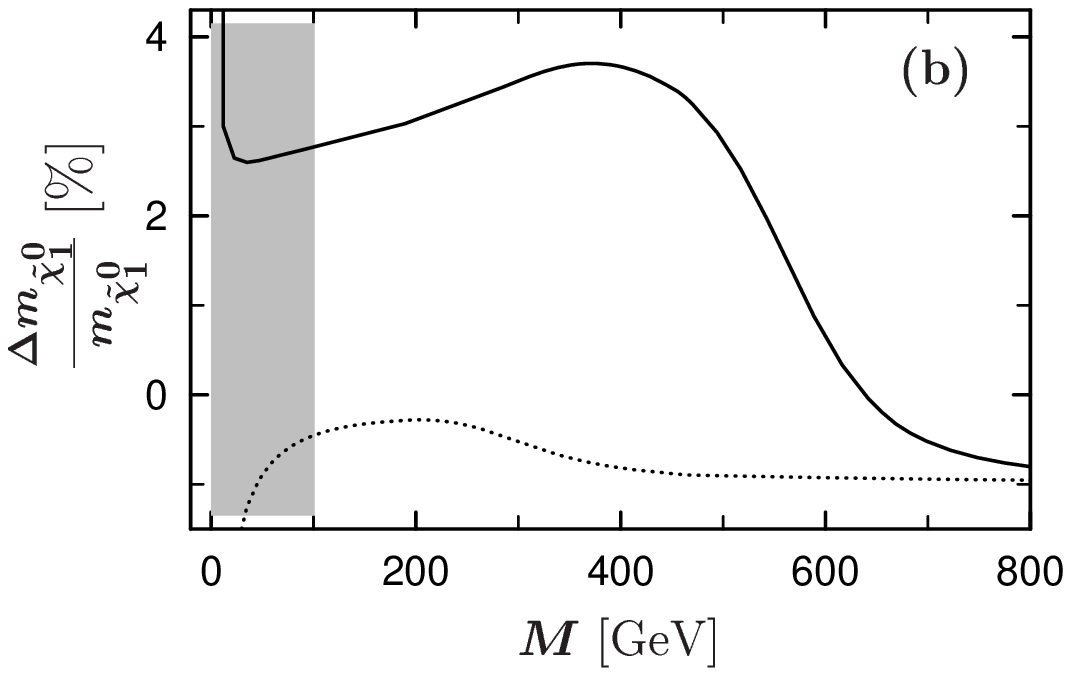}}}
 \hspace{-10mm}
 \vspace{-5mm}
 \caption[fig5]
 {Comparison of relative corrections to $m_{\chi^0_3}$ (a)
 $m_{\chi^0_1}$ (b). The full line shows the case where the SUSY
 SU(5) GUT relation is assumed for the \drbar parameters $M$ and $M'$,
 and the on--shell $M'$ is determined from $M$ by the same relation.
 The dotted line corresponds to the case where the on--shell $M'$
 is an independent parameter but satisfies the SUSY GUT relation.
 Other parameters are
$\tan{\beta}=7$, $\{M_{\tilde Q_1},\,M_{\tilde Q},\,A\}=
\{300,300,-500\}$~GeV, and $\mu = (-110\, {\rm (a)},\,-300\,{\rm
(b)})$~GeV. The grey areas are excluded by the bound
$m_{\ch^\pm_1}\geq\nolinebreak100$~GeV. }
 \label{fig:UnifVsNoUni}
 \end{center}
 \vspace{-7mm}
 \end{figure}

 Finally, we discuss the interesting case where $M'$
and $M$ are related by the SUSY SU(5) GUT relation
$M'=\frac{5}{3}\tan^2\theta_WM$ in the \drbar scheme. Then it is
possible to define the on--shell $M'$ by imposing this relation on
the on--shell parameters. In this case $\Delta Y_{11}$ gets a
contribution according to eq.~(\ref{eq:DeltaY11Uni}). This
correction is relatively large. For instance, for the parameter
set of Fig.~\ref{fig:mudep}a with $\mu=-130$~GeV one gets $\Delta
Y_{11}/Y_{11}\simeq 5.7/149$. This corresponds to the SUSY
threshold correction to the unification condition of the gaugino
masses \cite{pierce,kribs}. $\Delta Y_{11}$ gives a large
correction to the mass of a bino--like neutralino. This is clearly
seen in Fig.~\ref{fig:UnifVsNoUni}. Fig.~\ref{fig:UnifVsNoUni}a
shows the correction to $m_{\tilde{\chi}^0_3}$ as a function of
$M$ for $\mu=-130$~GeV and the other parameters as in
Fig.~\ref{fig:mudep}a. The solid line shows the case where the
\drbar parameters $M$ and $M'$ satisfy the SUSY GUT relation and
the on--shell $M'$ is defined by the same relation. For
comparison, the dotted line shows the case where the on--shell
$M'$ is defined by $Y_{11}$ as an independent parameter (with
$\Delta Y_{11}=0$), but its value coincides with the on--shell
$\frac{5}{3}\tan^2\theta_WM$. $\tilde{\chi}^0_3$ is bino--like for
$M>200$~GeV and $\delta m_{\tilde\chi^0_3}/m_{\tilde\chi^0_3}$
goes up to $4\%$. In Fig.~\ref{fig:UnifVsNoUni}b we show $\delta
m_{\tilde\chi^0_1}/m_{\tilde\chi^0_1}$ for $\mu=-300$~GeV. Here
$\tilde\chi^0_1$ is almost purely bino--like, hence it also gets a
large correction.

\section{Conclusions }
\label{sec:concl}

We have presented a consistent method for the calculation of the
one--loop corrections to the on--shell mass matrices of charginos
and neutralinos, and hence their masses. The on--shell parameters
$M$, $\mu$, and $M'$ are determined by the elements of the
on--shell mass matrices. We have calculated the corrections to the
tree--level mass matrices in terms of on--shell parameters. We
have performed a detailed numerical analysis of the corrections
due to fermion and sfermion loops, as function of the SUSY
parameters. When the parameters $M$ and $\mu$ are determined by
the chargino system, one gets corrections to the neutralino masses
of up to 4~\%. We have also treated the case where the on--shell
$M'$ is defined by $M$ using the SUSY GUT relation. Therefore,
these corrections have to be taken into account in precision
experiments at future $e^+ e^-$ linear colliders.

\section*{Acknowledgements}

The work of Y.\,Y. was supported in part by the Grant--in--aid for
Scientific Research from Japan Society for the Promotion of
Science, No.~12740131. H.~E., M.~K., and W.~M. thank the ''Fonds
zur F\"orderung der wissenschaftlichen Forschung of Austria'',
project no. P13139-PHY for financial support.

\begin{appendix}
\section*{Appendix}
\label{sec:app} \setcounter{equation}{0}
\renewcommand{\theequation}{A.\arabic{equation}}

In the following we give the formulas for the various
self--energies due to fermion and sfermion one--loop contributions
and the formulas for the counter terms of $m_Z$, $m_W$,
$\sin\theta_W$, and $\tan\beta$ used in this work.

\subsection*{Chargino self--energies}
\label{sec:app:charg}

The chargino self--energies read:
 \bea
 {\Pi}^L_{ij}(k^2)&=&
 -\,\forpi\,\Sgen\,N_C\,\Sa
  \bigg[l^\su_{ai}l^\su_{aj}\,B_1(k^2,m_d^2,m_{\su_a}^2)
 +k^\sd_{ai}k^\sd_{aj}\,B_1(k^2,m_u^2,m_{\sd_a}^2)\bigg]\,,
 \nonumber\\
 {\Pi}^R_{ij}(k^2)&=&
 -\,\forpi\,\Sgen\,N_C\,\Sa
  \bigg[k^\su_{ai}k^\su_{aj}\,B_1(k^2,m_d^2,m_{\su_a}^2)
 +l^\sd_{ai}l^\sd_{aj}\,B_1(k^2,m_u^2,m_{\sd_a}^2)\bigg]\,,
 \nonumber\\
 {\Pi}^{S,L}_{ij}(k^2)&=& \hspace{-0.6pt}
 \hphantom{-}\,\forpi\,\Sgen\,N_C\,\Sa
 \bigg[m_u\,l^\sd_{ai}k^\sd_{aj}\,B_0(k^2,m_u^2,m_{\sd_a}^2)
 +m_d\,k^\su_{ai}l^\su_{aj}\,B_0(k^2,m_d^2,m_{\su_a}^2)\bigg]\,,
  \nonumber\\
 {\Pi}^{S,R}_{ij}(k^2)&=&
 \hphantom{-}\,\forpi\,\Sgen\,N_C\,\Sa
 \bigg[m_u\,l^\sd_{aj}k^\sd_{ai}\,B_0(k^2,m_u^2,m_{\sd_a}^2)
 +m_d\,k^\su_{aj}l^\su_{ai}\,B_0(k^2,m_d^2,m_{\su_a}^2)\bigg]
 \,.\nonumber\\
&& \label{eq:chargSE}
 \eea
Here and in the following the index $u$ ($d$) denotes an up
(down)--type fermion, $\Sgen$ denotes the sum over all 6~fermion
generations. $N_C=3$ (1) in the quark (lepton) case. The
chargino--sfermion--fermion couplings are
 \bea
 l^\su_{ak}\;=\;-\,g\,V_{k1}R^{\su}_{a1}\,+\,
 h_u\,V_{k2}R^\su_{a1}\,,
 \hspace{30mm}
 k^\su_{ak}\;=\; h_d\,U_{k2}R^\su_{a1}\,,
 \nonumber\\
 l^\sd_{ak}\;=\;-\,g\,U_{k1}R^{\sd}_{a1}\,+\,
 h_d\,U_{k2}R^\sd_{a1}\,,
 \hspace{30mm}
 k^\sd_{ak}\;=\; h_u\,V_{k2}R^\sd_{a1}\,,
 \label{eq:couplk}
 \eea
where $U$, $V$ ($R^\sf$) are the chargino (sfermion) mixing
matrices and
 \bea
 h_u\;=\;\frac{g\;m_u}{\sqrt{2}\,m_W\,\sin{\beta}}\,,
 \hspace{20mm}
 h_d\;=\;\frac{g\;m_d}{\sqrt{2}\,m_W\,\cos{\beta}}\,.
 \label{eq:couplYuk}
 \eea
The two--point functions $B_0$ and $B_1$ \cite{thooft} are given
in the convention \cite{denner}.

\subsection*{Neutralino self--energies}
\label{sec:app:neu}

The neutralino self--energies read:
 \begin{eqnarray}
 \label{eq:neuSE}
 \Pi^{\ch^0\;L}_{ij}(k^2)\;\,=\;\,\, \Pi^{\ch^0\;R}_{ij}(k^2)\hspace{-2mm}
 &=&\!\!\!
  \frac{-1}{(4\pi)^2}\,\Sgen\,N_C\!
 \Sf \; \Sa \;(a^{\sf}_{ai} a^{\sf}_{aj}+
 b^{\sf}_{ai} b^{\sf}_{aj})\,
 B_1(k^2,m_f^2,m_{\sf_a}^2)\, ,\hspace{0.8cm}
 \\
 \label{eq:neuSES}
 \Pi^{\ch^0\,S,L}_{ij}(k^2) = \Pi^{\ch^0\,S,R}_{ij}(k^2)\hspace{-2mm} &=&
 \!\!\!
 \frac{1}{(4\pi)^2}\,\Sgen\,N_C\!
  \Sf \; \Sa \;(a^{\sf}_{ai}b^{\sf}_{aj}+
 a^{\sf}_{aj} b^{\sf}_{ai})\,
 B_0(k^2,m_f^2,m_{\sf_a}^2)\, .\hspace{0.8cm}
 \end{eqnarray}
The neutralino--sfermion--fermion couplings are
 \beq
  \label{eq:coupab}
 a^{\sf}_{ak} \,=\,
 g f_{Lk}^f R^\sf_{a1} +
 h_f Z_{kx} R^\sf_{a2}\,,
 \hspace{20mm}
 b^{\sf}_{ak} \,=\,
 g f_{Rk}^f R^\sf_{a2} +
 h_f Z_{kx}R^\sf_{a1}\,,
 \eeq
with $x=3$ for down--type and $x=4$ for up--type fermions. $Z$
denotes the neutralino mixing matrix and the terms $f_{Lk}^f$ and
$f_{Rk}^f$ are
 \bea
  \label{eq:fLi}
 f_{Lk}^f &=& \sqrt2\, \bigg[ \Big(e_f-\IL\Big) \tw Z_{k1}+\IL
 Z_{k2}\bigg]\,,
 \\
 \label{eq:fRi}
 f_{Rk}^f &=& -\sqrt2\, e_f \,\tw\, Z_{k1}\,.
  \eea

\subsection*{Gauge boson self--energies}
\label{sec:app:EW}

The counter terms for $m_W$ and $m_Z$ are given by
\begin{equation}
  \delta m^2_V = {\rm Re}\,  \Pi^{V V}_T(m^2_V), \qquad (V = W,
  Z)\, .
\label{eq:dmV}
\end{equation}
For the weak mixing angle we use the definition $\sin^2\theta_W =
1 - m^2_W/m^2_Z$. This gives \cite{sirlin}
\begin{equation}
  \frac{\delta \sin\theta_W}{\sin\theta_W} = \frac{1}{2}\frac{\cos^2\theta_W}{\sin^2\theta_W}\,
  \left( \frac{\delta m^2_Z}{m^2_Z} -
  \frac{\delta m^2_W}{m^2_W}\right)\, .
\label{eq:dsintW}
\end{equation}
The explicit forms for the self--energies are:
 \begin{eqnarray}
 \Pi^{WW}_T(k^2)
 &=&
 \forpi\,
 \gsh\,\Sgen N_C\Bigg[\Big(k^2-m_u^2-m_d^2\Big)B_0(k^2,m_u^2,m_d^2)\;+
 \;4B_{00}(k^2,m_u^2,m_d^2)
 \nonumber\\&&\hspace{30mm}
 -\;
  \Sf A_0(m_f^2)
  \;+\; \Sf\;\Sa
 \left(R^{\tilde{f}}_{a1}\right)^2 A_0(m_{\sf_a}^2)
 \nonumber
 \\*&&\hspace{30mm}
 -\;4\Sab\;\Big(R^{\su}_{a1}\Big)^2
 \left(R^{\sd}_{b1}\right)^2 B_{00}(k^2,m_{\su_a}^2,m_{\sd_b}^2)
  \;\;\Bigg]\,,\label{eq:Wself}\\[5mm]
 \Pi^{ZZ}_T(k^2)
  &=&
 \forpi\,
 \left(\frac{g}{\cos{\theta_W}}\right)^2\Sgen N_C\,\Sf
 \Bigg\{
 2\Sa\,\left[\CLs\Big(R^\sf_{a1}\Big)^2+
 \CRs\Big(R^\sf_{a2}\Big)^2\right]A_0(m_{\sf_a}^2)
 \nonumber\\*&&\hspace{20mm}-\;
 4\Sab
 \left[C_L^fR^\sf_{a1}R^\sf_{b1}+
 C_R^fR^\sf_{a2}R^\sf_{b2}\right]^2
 B_{00}(k^2,m_{\sf_a}^2,m_{\sf_b}^2)
 \nonumber\\*&&\hspace{20mm}-\;
 \left[\CLs+\CRs\right]2\bigg[A_0(m_f^2)-2B_{00}(k^2,m_f^2,m_f^2)\bigg]
 \nonumber\\*&&\hspace{7mm}+\;
 \bigg[k^2\left(\CLs+\CRs\right)-
 2m_f^2\left(\CL-\CR\right)^2\bigg]B_0(k^2,m_f^2,m_f^2)
 \Bigg\}\,,\label{eq:Zself}
 \eea
with
 \beq
 \CL\,=\,\IL-\sin^2{\theta_W}\, e_f\,,
 \hspace{25mm}
 \CR\,=\,-\sin^2{\theta_W}\, e_f\,.
 \label{eq:CLR}
 \eeq
The functions $A_0$, $B_0$, $B_{00}$, \cite{thooft} are given in
the convention of \cite{denner}.

\subsection*{\boldmath $A^0$--$Z^0$ self--energy}
\label{sec:app:AZ}

The mixing angle $\beta$ is fixed by the condition
\cite{pokorski}:
 \beq
 \textrm{Im}\,\hat{\Pi}_{A^0Z^0}(m_A^2)\,=\,0\,.
 \label{eq:AZSEfixing}
 \eeq

The renormalized  self--energy $\hat{\Pi}_{A^0Z^0}(k^2)$ is
defined by the two--point function\\ {\setlength{\unitlength}{1mm}
\begin{picture}(160,23)(0,0)
\put(0,-2){\mbox{\resizebox{8cm}{!}{\includegraphics{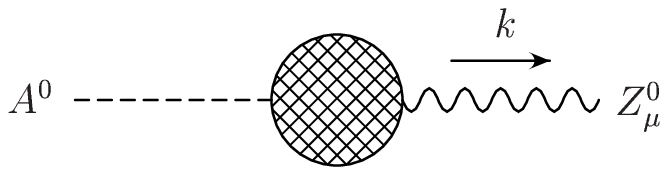}}}}
\put(75,9){\makebox(0,0)[l]{
  $=\,\,-\,i\,k^\mu\,\hat{\Pi}_{A^0Z^0}(k^2)\,\epsilon^*_\mu
  (k)\,.$}}
\end{picture}}

Thus we get the counter term for $\tan\beta$:
 \beq
 \frac{\d\tan{\beta}}{\tan{\beta}}\;=\;
- \frac{1\;}{m_Z\,\sin{2\beta}}\;
 {\textrm{Im}\,}{\Pi}_{A^0Z^0}(m_A^2)\,.
 \label{eq:dtanb}
 \eeq
${\Pi}_{A^0Z^0}$ denotes the unrenormalized self--energy
 \bea
\hspace{-10mm}
 {\Pi}_{A^0Z^0}(k^2)&=&
 \frac{i}{\left(4\pi\right)^2}\;m_Z\,\sin{2\beta}\,\Sgen\,N_C\,\Sf\,\IL\,h_f^2\,
 \Bigg\{B_0(k^2,m_f^2,m_f^2)
 \nonumber\\&&\hspace{5mm}
 +\;
 \frac{\;\sin{2\theta_\sf}\;}{2m_f}\bigg[A_f+\mu
 \left(\begin{array}{@{\,}c@{\,}}\tan{\beta} \\ \cot{\beta}\,
 \end{array}\right)
 \bigg]\;\Big(2B_1+B_0\Big)(k^2,m_{\sf_1}^2,m_{\sf_2}^2)\Bigg\}\,.
 \label{eq:A0Z0self}
 \eea

\end{appendix}

\end{document}